\documentclass[journal]{IEEEtran}

\usepackage{cite}
\usepackage{amsmath,amssymb,amsfonts}
\usepackage{graphicx}
\usepackage{algorithm,algorithmic}
\usepackage{booktabs,multirow,threeparttable}
\usepackage{url}
\usepackage{textcomp}
\usepackage[hidelinks]{hyperref}

\begin{document}

\title{Learning from Limited Multi-Phase CT: Dual-Branch Prototype-Guided Framework for Early Recurrence Prediction in HCC}

\author{%
Hsin-Pei Yu$^{1}$, Si-Qin Lyu$^{2}$, Yi-Hsien Hsieh$^{1}$, Weichung Wang$^{3}$,\\
Tung-Hung Su$^{4}$, Jia-Horng Kao$^{4}$, and Che Lin$^{1,2,5,6,*}$%
\thanks{$^{1}$ Graduate Institute of Communication Engineering, National Taiwan University (NTU), Taipei 10617, Taiwan.}%
\thanks{$^{2}$ Smart Medicine and Health Informatics Program, NTU, Taipei 10617, Taiwan.}%
\thanks{$^{3}$ Institute of Applied Mathematical Sciences, NTU, Taipei 10617, Taiwan.}%
\thanks{$^{4}$ Division of Gastroenterology and Hepatology, Department of Internal Medicine, National Taiwan University Hospital, Taipei 10048, Taiwan.}%
\thanks{$^{5}$ Department of Electrical Engineering, NTU, Taipei 10617, Taiwan.}%
\thanks{$^{6}$ Center for Advanced Computing and Imaging in Biomedicine, NTU, Taipei 10617, Taiwan.}%
\thanks{*Corresponding author: chelin@ntu.edu.tw}%
}

\maketitle
\thispagestyle{plain}

\begin{abstract}

Early recurrence (ER) prediction after curative-intent resection remains a critical challenge in the clinical management of hepatocellular carcinoma (HCC). 
Although contrast-enhanced computed tomography (CT) with full multi-phase acquisition is recommended in clinical guidelines and routinely performed in many tertiary centers, complete phase coverage is not consistently available across all institutions. In practice, single-phase portal venous (PV) scans are often used alone, particularly in settings with limited imaging resources, variations in acquisition protocols, or patient-related factors such as contrast intolerance or motion artifacts.
This variability results in a mismatch between idealized model assumptions and the practical constraints of real-world deployment, underscoring the need for methods that can effectively leverage limited multi-phase data.
To address this challenge, we propose a Dual-Branch Prototype-guided (DuoProto) framework that enhances ER prediction from single-phase CT by leveraging limited multi-phase data during training. DuoProto employs a dual-branch architecture: the main branch processes single-phase images, while the auxiliary branch utilizes available multi-phase scans to guide representation learning via cross-domain prototype alignment. Structured prototype representations serve as class anchors to improve feature discrimination, and a ranking-based supervision mechanism incorporates clinically relevant recurrence risk factors. Extensive experiments demonstrate that DuoProto outperforms existing methods, particularly under class imbalance and missing-phase conditions. Ablation studies further validate the effectiveness of the dual-branch, prototype-guided design. Our framework aligns with current clinical application needs and provides a general solution for recurrence risk prediction in HCC, supporting more informed decision-making. 

\end{abstract}


\vspace{-0.4cm}

\section{Introduction}
\label{sec:introduction}
\IEEEPARstart{H}{epatocellular Carcinoma} (HCC) is one of the leading causes of cancer-related mortality worldwide, with high recurrence rates even after curative-intent hepatic resection\cite{bray2024global}. Among these, early recurrence (ER) refers to recurrence that occurs within two years after surgery and poses a particularly severe threat to long-term survival, while patients without recurrence during this period are considered non-early recurrence (NER). Over 80\% of HCC patients experience recurrence within this timeframe \cite{pagano2022hepatocellular}, underscoring the urgent need for accurate ER prediction. A reliable ER prediction model can guide surgical planning and adjuvant therapy, ultimately improving clinical outcomes.

\begin{figure}[!t]
\includegraphics[width=\linewidth]{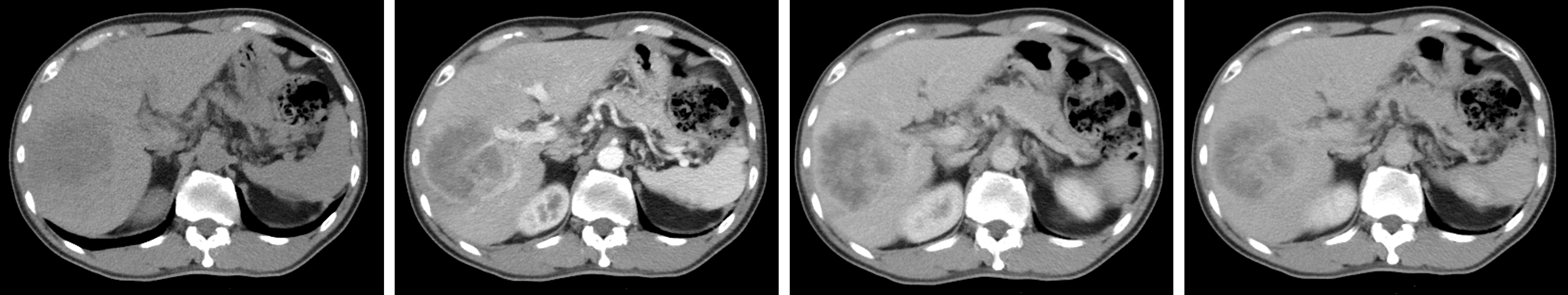}
\caption{Multi-phase CT scan illustrating intensity changes. From left to right: pre-contrast (Pre), arterial (A), portal venous (PV), and delayed (D) phases.}
\label{fig2}
\vspace{-0.52cm}

\end{figure}

Contrast-enhanced computed tomography (CT) is widely used for preoperative evaluation in HCC patients. A standard imaging protocol includes a pre-contrast (Pre) scan followed by three post-contrast phases: arterial (A), portal venous (PV), and delayed (D). These multi-phase scans capture how contrast moves through the liver and tumor tissue over time. For example, HCC lesions become brighter during the A phase and darker in the PV or D phases  (Fig.~\ref{fig2}). This pattern provides important clues about tumor characteristics and diagnosis. While multi-phase imaging offers complementary temporal information, complete acquisition of all phases is often infeasible in real-world clinical settings\cite{uhm2022unified}. Incomplete phase coverage may result from missing data, patient intolerance to contrast agents, abbreviated scan protocols in emergency or resource-limited settings, or institutional workflow differences.

Despite recent advances in deep learning for HCC prognosis modeling, most existing methods focus exclusively on either multi-phase or single-phase CT scans \cite{zhang2024multimodal, wang2022PhaseAttentionModel, song2024EarlyRecurrencePrediction}. Few studies consider both modalities \cite{wang2023prototype}, and even fewer explore how limited multi-phase data can be used to support learning from the more prevalent single-phase scans---an approach that better reflects real-world clinical application needs. In particular, prior work has not fully addressed the challenge of effectively aligning and transferring complementary information across heterogeneous imaging phases.

To bridge this gap, we introduce a dual-branch framework that jointly leverages both single-phase and limited multi-phase CT scans for ER prediction. The main branch focuses on the clinically prevalent single-phase inputs and is trained to predict the primary recurrence risk target. To enhance its learning, we incorporate an auxiliary branch that utilizes available multi-phase data, which provides richer semantic context and phase-specific enhancement cues. To facilitate meaningful interaction between branches, we propose a prototype-guided mechanism that constructs and matches class-specific prototypes---feature space centroids---from both branches. These prototypes act as anchors within the feature maps, enabling representation alignment between the heterogeneous imaging inputs. This prototype-guided learning captures class-discriminative patterns that generalize across patients and imaging protocols, thereby improving training efficiency and model robustness. We refer to our framework as Dual-Branch Prototype-guided (DuoProto) learning, which promotes structured feature alignment and discriminative representation learning under modality heterogeneity.

Our main contributions are summarized as follows:
\begin{itemize}
    \item We propose a unified dual-branch learning framework that integrates limited multi-phase CT data with widely available single-phase scans to improve ER prediction in HCC.

    \item We introduce a prototype-guided learning mechanism that provides stable, class-discriminative representations and facilitates effective knowledge sharing across patients and imaging domains.

    \item Our design is designed to reflect real-world clinical application needs and improve model performance under limited multi-phase supervision, offering a practical solution for clinical deployment.
\end{itemize}

\vspace{-0.2cm}
\section{Related Works}
\label{sec:related_work}

\subsection{Early Recurrence Prediction for HCC}
Early recurrence (ER) remains a critical factor affecting long-term survival in HCC, and has therefore drawn increasing attention in both clinical and research communities. Radiomics-based models extract handcrafted features from CT or MRI, but they often rely on manual feature design and are sensitive to variations in acquisition protocols and patient populations \cite{jin2024radiomics, radiomics}. Several studies further combined imaging with clinical variables to enhance prognostic\cite{hui2024MultimodalMultiphasicPreoperative, kinoshita2023DeepLearningModel}, yet clinical data are frequently missing or inconsistently recorded in real-world practice. 

Beyond clinical integration, several studies have further explored the use of multi-phase CT imaging to capture tumor perfusion dynamics\cite{zhang2024multimodal}. Wang et al. proposed a phase attention model to adaptively fuse information from multi-phases scans\cite{wang2022PhaseAttentionModel}, while Song et al. introduced self-supervised pretraining tasks to extract intra- and inter-phase representations\cite{song2024EarlyRecurrencePrediction}. Although these methods achieved promising results, they typically rely on substantial amounts of multi-phase inputs, which are not always available in practice due to real-world acquisition constraints. In contrast, our method unifies single- and multi-phase representations via prototype-guided training, enabling robust recurrence prediction despite incomplete or heterogeneous imaging phases.

\vspace{-0.2cm}
\subsection{Prototype-Based Representation Learning}
Prototype learning was first introduced in few-shot learning to represent each class by a feature centroid in the embedding space, enabling efficient and interpretable metric-based inference \cite{snell2017prototypical}. Since then, prototype-based methods have been widely adopted across a range of tasks, including semantic segmentation, long-tailed classification, and medical image analysis, where class-specific anchors stabilize learning and enhance feature discrimination under limited supervision \cite{wang2019panet, yang2022proco}. In the medical domain, prototypes have been leveraged to improve feature consistency and interpretability\cite{liu2024pamil, cheng2023prior, HOPE}.
Inspired by Wang et al.’s prototype knowledge distillation framework \cite{wang2023prototype}, we adopt prototype supervision not for distillation, but for learning structured representations across phases. Specifically, prototypes in our model serve as semantic anchors to facilitate alignment and discriminability in heterogeneous and modality-scarce settings.

\vspace{-0.2cm}
\subsection{Learning with Missing Phase}
Missing phase is a common challenge in real-world medical imaging, where certain imaging phases are absent due to acquisition constraints. Prior studies have explored strategies such as modality synthesis, latent space alignment, and generative adversarial networks \cite{azad2022medical}. However, these approaches often suffer from noisy reconstructions, high computational costs, or weak supervision.

More recently, knowledge distillation (KD) has emerged as a lightweight yet effective approach, where a model trained on complete modalities transfers information to one operating under incomplete inputs. Among these, prototype-guided distillation shows promise in aligning latent features across modalities. For example, ProtoKD utilizes a multi-phase teacher to guide a single-phase student via prototype alignment and feature-level supervision \cite{wang2023prototype}.

However, such designs are based on a strict teacher-student formulation that assumes abundant fully paired multi-phase data. In contrast, our approach adopts a different perspective, more reflective of real-world datasets, where multi-phase scans are often incomplete due to workflow constraints. Rather than enforcing hierarchical supervision, we treat multi-phase CT as an auxiliary modality and jointly train both branches under prototype-guided alignment. This enables effective knowledge sharing despite data heterogeneity and label sparsity.

\vspace{-0.2cm}
\section{Methods}

\label{sec:method}

\subsection{Problem Formulation}
We aim to perform binary ER prediction in HCC patients using contrast-enhanced 3D CT. Given a liver volume $X \in \mathbb{R}^{H \times W \times D}$, where $H$, $W$, and $D$ represent the height, width, and depth of the volume, respectively. The model outputs an uncalibrated probability $\Tilde{y} \in [0, 1]$ indicating the likelihood of recurrence within a predefined postoperative interval. A binary label $\hat{y}_i =1$ is assigned to patient $i$ if its predicted likelihood of recurrence is above a specified threshold, and $\hat{y}_i =0$ is assigned otherwise.

Regarding the model input, we leverage both single/multi-phase information sources. Every patient possesses a single PV-phase CT scan, represented as $\mathbf{x}_i^{\text{s}} = \{ x_i^{\text{PV}} \}$, given its diagnostic importance and routine use in clinical workflows. On the other hand, a limited subset of patients contains available multi-phase (Pre, A, PV, and D) CT scans, denoted as $\mathbf{x}_i^{\text{m}} = \{ x_i^{\text{Pre}}, x_i^{\text{A}}, x_i^{\text{PV}}, x_i^{\text{D}} \}$, which capture temporally diverse contrast phases and provide additional information. 

Our proposed model consists of two branches (multi-phase/single-phase), each parameterized by separate encoders and projection heads. During training, we minimize a joint objective function that includes multiple supervised losses (see Section~\ref{sec:proto-machanism}) to optimize both branches simultaneously. The multi-phase branch serves as an auxiliary signal during training to help refine the representation space, while the single-phase branch serves as the main predictor at inference time. A binary prediction is obtained by thresholding the predicted probability. For more details about thresholding, please see Section~\ref{sec:eval}.

\vspace{-0.2cm}
\subsection{Overview Architecture}
\begin{figure*}[!t]
  \centering
  \includegraphics[width=\linewidth]{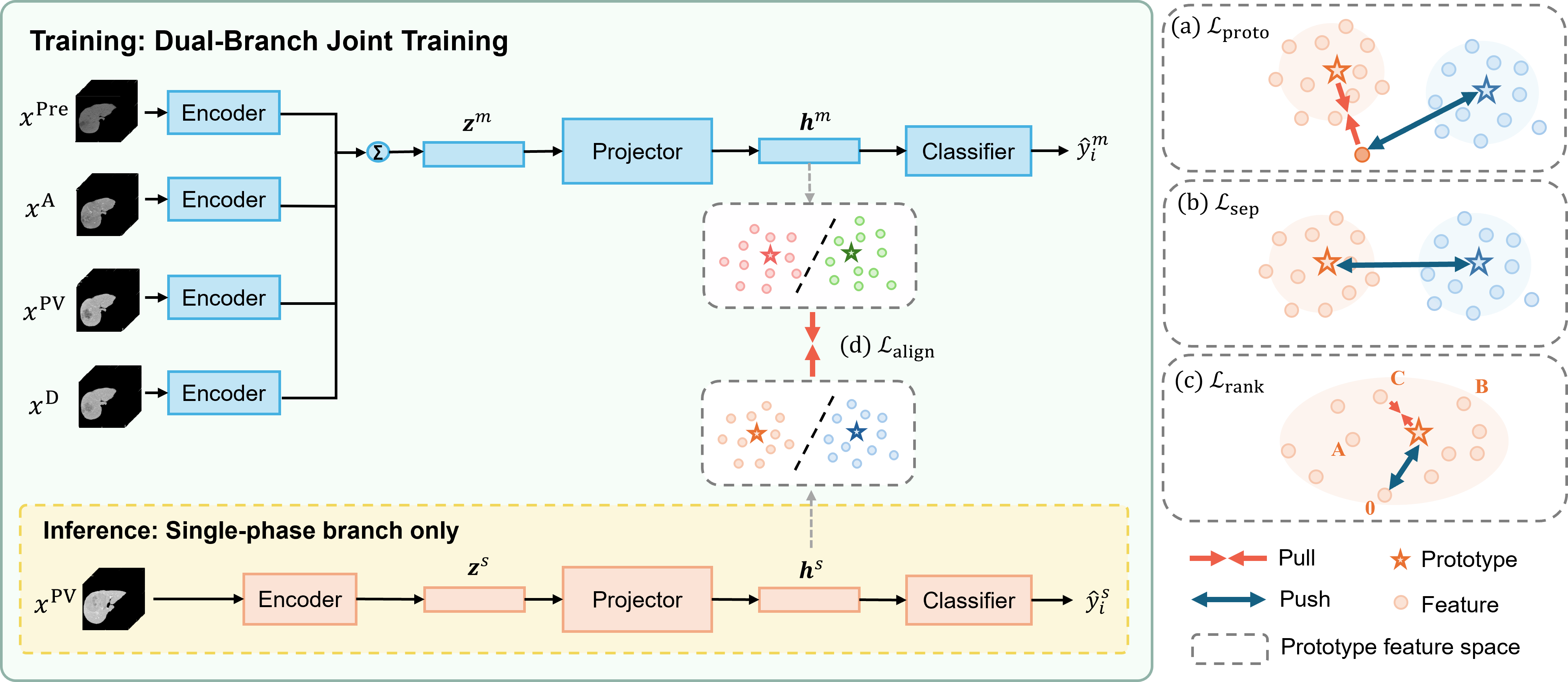}
  \caption{Overall framework of the proposed dual-branch (upper: multi-phase; lower: single-phase) prototype-guided model. (a)~Contrastive prototype loss ($\mathcal{L}_{\text{proto}}$) pulls features toward their class-specific prototypes. (b)~Prototype separation loss ($\mathcal{L}_{\text{sep}}$) increases margins between class prototypes. (c)~BCLC-informed ranking loss ($\mathcal{L}_{\text{rank}}$) encourages prototype distances to reflect ordinal staging. (d)~Prototype alignment loss ($\mathcal{L}_{\text{align}}$) aligns single-phase and multi-phase prototypes across domains. For each branch, the encoder output $z$ is projected to $h$ in the prototype feature space for further calculation. We use superscripts $m$ and $s$ to denote features from the multi-phase and single-phase branches, respectively.}

  \label{fig:overall}
\vspace{-0.51cm}
\end{figure*}

As shown in Fig.~\ref{fig:overall}, DuoProto adopts a dual-branch architecture to jointly process multi-phase and single-phase CT inputs. 

We adopt the ResNet-ViT hybrid backbone from our previous work~\cite{ReViT}, which is tailored for volumetric data and integrates local detail extraction with global context modeling. Each volumetric input is partitioned into 3D patches and encoded via modality-specific encoders based on this backbone.
In the multi-phase branch, each phase is processed independently, and the resulting features are integrated via a late attention fusion module to form a phase-aware representation. In the single-phase branch, the PV phase is encoded separately to obtain its corresponding feature representation.

Let $\mathbf{z}_i = f(\mathbf{x}_i)$ denote the volumetric-level feature obtained by average pooling the encoded patch-level features from input $\mathbf{x}_i$. This feature vector is then mapped into a prototype space through a projection head  $g(\cdot)$ consisting of two fully connected layers with batch normalization and ReLU activation, producing $\mathbf{h}_i = g(\mathbf{z}_i) \in \mathbb{R}^d$, where $d$ denotes the hidden state dimension. The projected feature $\mathbf{h}_i$ is $\ell_2$-normalized and used for both classification and prototype-guided supervision. A final linear classifier produces logits for ER prediction, while a shared prototype set maintains class-specific prototypes for alignment between the two branches, which will be detailed in subsection~\ref{sec:proto-machanism}.

Unlike the teacher-student paradigm in \cite{wang2023prototype}, our dual-branch framework treats multi-phase inputs as auxiliary modality during training. By employing prototype-guided alignment, the model encourages semantic consistency across modalities, enabling effective representation learning. At inference, only the single-phase branch is used, ensuring compatibility with deployment scenarios where PV-phase CT is commonly retained or preferred in practice.

\vspace{-0.2cm}
\subsection{Prototype-guided Mechanism}
\label{sec:proto-machanism}

Under conditions of heterogeneous data distributions and limited sample sizes, direct supervision using cross-entropy loss often struggles to produce stable and semantically structured feature representations. This challenge is especially pronounced in ER prediction for HCC, where subtle imaging variations and class imbalance introduce difficulties for discriminative learning. To address this, we propose a prototype-guided learning mechanism in which each ER class is associated with a latent prototype vector that serves as a geometric anchor in the feature space.

These prototypes are iteratively updated during training using aggregated hidden states across patients and serve two primary roles: (1) aligning semantic structures between the single-phase and multi-phase domains, and (2) functioning as shared representations at the population level that can be used for regulating intra-class consistency and inter-class separation. This design enables the model to leverage both cross-modality alignment and cross-patient regularizers, while maintaining compatibility with single-phase inputs at inference time.

\subsubsection{Prototype Initialization and Update Strategy}
We associate each ER class $c \in \{0, 1\}$ with a prototype vector $\mathbf{p}_c \in \mathbb{R}^d$, which acts as a class-level semantic anchor in the latent space. To ensure numerical stability and semantic consistency throughout training, we adopt a hybrid update strategy consisting of batch-based initialization and exponential moving average (EMA) updates.

At the first iteration \(t=0\) in which samples from class \(c\) appear, the prototype is initialized using the centroid of the projected features from that class in the mini-batch: 
\begin{equation}
\mathbf{p}_c^{(0)} = \frac{1}{|\mathcal{B}_c^{(0)}|} \sum_{i \in \mathcal{B}_c^{(0)}} \mathbf{h}_i,
\label{eq:init_proto}
\end{equation}
where $\mathcal{B}_c^{(0)} = \{ i \mid y_i = c \}$  denotes the set of sample indices with ground-truth label \( c \), and $\mathbf{h}_i \in \mathbb{R}^d$ is the normalized projected feature of instance \( i \).

In all subsequent iterations \(t>0\), prototypes are updated using an EMA rule with fixed momentum $\mu = 0.9$:
\begin{equation}
\mathbf{p}^{(t)}_c = \mu \cdot \mathbf{p}^{(t-1)}_c + (1 - \mu) \cdot \overline{\mathbf{h}}^{(t)}_c.
\label{eq:proto_update}
\end{equation}
Note that the batch-level centroid feature $\overline{\mathbf{h}}_c^{(t)}$ is computed as:

\begin{equation}
\overline{\mathbf{h}}_c^{(t)} = \frac{1}{|\mathcal{B}_c^{(t)}|} \sum_{i \in \mathcal{B}_c^{(t)}} \mathbf{h}_i,
\label{eq:batch_mean}
\end{equation}
where $\mathcal{B}_c^{(t)}$ denotes the set of indices for samples from class $c$ in the current mini-batch at iteration $t$. All feature vectors $\mathbf{h}_i$ and prototypes $\mathbf{p}_c$ are $\ell_2$-normalized to maintain angular consistency in the prototype space. (Details in Supplementary I.)

\subsubsection{Contrastive Prototype Loss}
To enforce intra-class consistency and encourage semantic alignment in the latent space, we employ a contrastive prototype loss that attracts each projected feature vector $\mathbf{h}_i \in \mathbb{R}^d$ toward its corresponding class prototype $\mathbf{p}_{y_i} \in \mathbb{R}^d$ while implicitly repelling it from the opposing class prototype. 
The contrastive prototype loss is formulated as:
\vspace{-0.1cm}
\begin{equation}
\mathcal{L}_{\text{proto}} = - \log \frac{\exp\left(\cos(\mathbf{h}_i, \mathbf{p}_{y_i}) / \tau\right)}{\sum_{c=0}^{1} \exp\left(\cos(\mathbf{h}_i, \mathbf{p}_c) / \tau\right)},
\label{eq:loss_proto}
\end{equation}
where $\tau > 0$ is a temperature parameter that controls the sharpness of the similarity distribution, and $\cos(\cdot, \cdot)$ denotes cosine similarity between unit-normalized vectors.

This loss guides the network to cluster features around their respective class prototypes, thereby improving intra-class consistency and enhancing the semantic structure of the embedding space. It also promotes better generalization by reducing the overlap between classes in the latent space and provides an interpretable geometric basis for feature-to-prototype attraction.

\subsubsection{Prototype Separation Loss}
While the contrastive prototype loss encourages intra-class consistency, it does not explicitly enforce separation between class prototypes. To prevent prototype collapse and ensure inter-class disentanglement, we introduce a prototype separation loss that imposes a minimum angular margin between the semantic anchors of different classes.

Let $\mathbf{p}_0$ and $\mathbf{p}_1 \in \mathbb{R}^d$ denote the normalized prototypes for NER and ER classes, respectively. Both vectors are constrained to lie on the unit hypersphere. The separation loss is defined as:
\vspace{-0.1cm}
\begin{equation}
\mathcal{L}_{\text{sep}} = \left[ \max(0, \cos(\mathbf{p}_0, \mathbf{p}_1) - m) \right]^2,
\label{eq:loss_sep}
\end{equation}
where $m \in [0, 1]$ is a predefined angular margin (set to $m = 0.8$), which we found to balance inter-class separation without destabilizing prototype updates.

This term penalizes the prototypes if their angular distance becomes too small, thereby encouraging them to remain geometrically distinct in the embedding space. By enforcing prototype-to-prototype separation, this loss enhances inter-class discriminability, improves decision boundary stability, and mitigates the effects of class overlap and label noise---particularly important in real-world clinical data with subtle recurrence patterns.

\subsubsection{BCLC-Informed Ranking Loss}
Distinguishing ER is challenging due to subtle imaging cues and class imbalance. To alleviate these issues, we incorporate clinically informed ordinal priors derived from the Barcelona Clinic Liver Cancer (BCLC) staging system, which provides ordinal estimates of tumor burden and progression risk. Consistent with clinical practice, our cohort also shows a significant association between BCLC stage and ER status ($p < 0.0001$), supporting its utility as an auxiliary ordinal prior.

Specifically, BCLC stages ($0$, $A$, $B$, $C$) reflect increasing tumor severity associated with ER risk. To inject this prior knowledge into the learning process, we design a weakly supervised ranking loss that applies to pairs of ER-positive patients. For any pair of patients $(i, j)$ such that $y_i = y_j = 1$ and $s_i > s_j$ (i.e., BCLC$_i$ $>$ BCLC$_j$), we enforce a soft ordering constraint:

\vspace{-0.2cm}
\begin{equation}
s_i > s_j \;\; \Rightarrow \;\; d(\mathbf{h}_i, \mathbf{p}_1) < d(\mathbf{h}_j, \mathbf{p}_1),
\label{eq:loss_proto}
\end{equation}
where $d(\cdot, \cdot)$ denotes a distance metric between feature vectors $\mathbf{h}$ and prototype $\mathbf{p}_1$ suggesting ER.

The ranking loss is then defined as:
\vspace{-0.1cm}
\begin{equation}
\mathcal{L}_{\text{rank}} = \frac{1}{|\mathcal{P}|} \sum_{(i, j) \in \mathcal{P}} \left[ \max\left( 0,\; d(\mathbf{h}_i, \mathbf{p}_1) - d(\mathbf{h}_j, \mathbf{p}_1) + \epsilon \right) \right]^2,
\label{eq:loss_rank}
\end{equation}
where $\mathcal{P} = \{(i,j) \mid s_i > s_j,\; y_i = y_j = 1 \}$ is the set of ER-positive patient pairs with ordered BCLC stages, and $\epsilon$ is a positive margin hyperparameter that controls the tolerance for ranking violations.

This loss introduces a fine-grained, relational structure into the embedding space, treating BCLC not as a discrete classification target but as a soft continuum of recurrence severity. By aligning feature distances with clinically meaningful ordering, the model is guided to better prioritize high-risk ER patients---thereby improving sensitivity and addressing borderline cases that are otherwise difficult to classify.

\subsubsection{Prototype Alignment Loss}
To align multi-phase and single-phase representations, we introduce a prototype alignment loss that enforces semantic consistency across modalities. Rather than aligning instance-level features---which may be modality-dependent and noisy---we align the class-wise prototypes learned from each branch. 

Let $\mathbf{p}_c^{\text{m}}$ and $\mathbf{p}_c^{\text{s}} \in \mathbb{R}^d$ denote the prototypes for class $c \in \{0,1\}$ learned from the multi-phase and single-phase branches, respectively, with superscripts $\text{m}$ and $\text{s}$ indicating multi-phase and single-phase modalities. Each prototype is normalized to unit length. The alignment loss is defined as the squared Euclidean distance between corresponding prototypes:
\vspace{-0.1cm}
\begin{equation}
\mathcal{L}_{\text{align}} = \sum_{c=0}^{1} \left\| \mathbf{p}_c^{\text{m}} - \mathbf{p}_c^{\text{s}} \right\|_2^2.
\label{eq:proto_align}
\end{equation}

This loss encourages both branches to converge toward a shared prototype space, promoting semantic-level alignment. Since prototypes encode global class-level representations, their alignment provides a more stable and modality-invariant learning signal. This design is particularly beneficial in scenarios where multi-phase CTs have limited sample size and serve only as an auxiliary modality during training, while PV-phase CTs remain the sole input during inference.

\subsubsection{Total Loss}

The overall training objective integrates the main ER classification loss with multiple auxiliary terms that guide representation learning. The total loss is defined as:
\vspace{-0.1cm}
\begin{equation}
\mathcal{L}_{\text{total}} = \alpha \cdot \mathcal{L}_{\text{CE}} + 
\beta \cdot \mathcal{L}_{\text{proto}} + 
\gamma \cdot \mathcal{L}_{\text{sep}} + 
\delta \cdot \mathcal{L}_{\text{rank}} + 
\lambda \cdot \mathcal{L}_{\text{align}},
\label{eq:total_loss}
\end{equation}
where $\mathcal{L}_{\text{CE}}$ is the primary cross-entropy loss for ER prediction, and the remaining terms correspond to the contrastive prototype loss ($\mathcal{L}_{\text{proto}}$), prototype separation loss ($\mathcal{L}_{\text{sep}}$), BCLC-informed ranking loss ($\mathcal{L}_{\text{rank}}$), and prototype alignment loss ($\mathcal{L}_{\text{align}}$), respectively.

We empirically set the weight coefficients as follows: $\alpha = 1.0$, $\beta = 0.6$, $\gamma = 0.8$, $\delta = 0.5$, and $\lambda = 0.1$. These values were selected based on validation performance and stability across experiments. The main classification objective is thus prioritized, while the auxiliary losses contribute to improving feature structure, semantic alignment, and clinical robustness without overwhelming the optimization process. (Training steps summarized in Supplementary IV.)

\vspace{-0.2cm}
\section{Experimental Setting}
\label{sec:experiment}
\subsection{Dataset}

We retrospectively collected data from patients diagnosed with HCC at National Taiwan University Hospital (NTUH) between August 2011 and August 2021. Inclusion criteria included: (1) age $\ge$ 18 years, (2) pathologically confirmed HCC, (3) curative-intent hepatectomy without prior treatment, and (4) contrast-enhanced CT performed within 3 months before surgery. Exclusion criteria comprised: (1) incomplete imaging data (i.e., CT scans that did not fully cover the liver region, or missing required contrast phases) and (2) follow-up duration less than 2 years postoperatively, which was necessary to reliably assess recurrence status. After applying the criteria, a total of 240 patients were included in the study.

ER was defined as recurrence detected within 2 years following surgery, while patients without recurrence during this period were considered NER. Patient demographics and recurrence distribution are summarized in Table~\ref{tab:demographics}.

Of the 240 patients included, 46 had complete multi-phase (Pre/A/PV/D) CT scans. The remaining 194 patients had only a single available phase, primarily the PV phase, which is commonly used in routine clinical workflows.

\vspace{-0.2cm}

\begin{table}[ht]
\centering
\caption{Patient demographics stratified by imaging phase availability.}
\label{tab:demographics}
\renewcommand{\arraystretch}{1.3}
\resizebox{\columnwidth}{!}{
\begin{tabular}{lcccc}
\toprule
\textbf{Variable} &  \textbf{Multi-phase} \newline (n=46) & \textbf{Single-phase} \newline (n=240) & \textbf{\textit{p}-value} \\
\midrule
Age (years) &  63.3 ± 10.9 & 61.8 ± 11.2 & 0.4041 \\
Gender (M/F) & 36 / 10 & 202 / 38 & 0.4434 \\
BCLC stage (0/A/B/C)  & 5 / 18 / 16 / 7 & 21 / 94 / 95 / 30 & 0.887 \\
Tumor size (cm)  & 6.62 ± 4.80 & 5.85 ± 4.29 & 0.2751 \\
\addlinespace[2pt]
\textbf{ER / NER (n, \%)} & \textbf{21 (45.7\%) / 25 (54.3\%)} & \textbf{90 (37.5\%) / 150 (62.5\%)} & \textbf{0.3820} \\
\bottomrule
\end{tabular}
}
\renewcommand{\arraystretch}{1.0}
\vspace{-0.4cm}
\end{table}
\vspace{-0.2cm}
\subsection{Data preprocessing}
We followed the preprocessing strategy described in \cite{ReViT}, as it offers a well-validated pipeline for standardizing 3D CT volumes in HCC studies. Given the similar imaging characteristics and anatomical variability between staging and recurrence tasks, we adopted and modified their approach to better suit ER prediction as follows.

All CT scans were resampled to a uniform voxel size of  \(1.0 \times 1.0 \times 1.0 \, \text{mm}^3\) to account for variable slice spacing across patients. Intensities were clipped to \([-21, 189]\) Hounsfield Units (HU) and normalized to [0, 1]. The liver region was segmented using a pretrain segmentation model, SegVol \cite{du2025segvoluniversalinteractivevolumetric}, a state-of-the-art (SOTA) segmentation model capable of accurately delineating over 200 anatomical structures from 3D medical images. We applied the resulting liver mask to each phase to remove irrelevant background and non-hepatic tissues.

To generate standardized inputs, we cropped a bounding box around the segmented liver and applied symmetric padding to reach a fixed size of \(192 \times 192 \times 192 \). If the cropped region exceeded this size, center cropping was applied. This design ensures spatial consistency while preserving anatomical structure.

We chose to retain the entire liver region of interest (ROI) instead of tumor-only regions because important recurrence-related features such as vascular invasion and liver morphology may exist outside tumor boundaries. Additionally, whole-liver segmentation is typically more reliable than tumor segmentation in clinical practice.

During training, we applied data augmentation including affine transformation, contrast adjustment, Gaussian noise, and dropout. These augmentations enhance generalizability and mitigate overfitting, especially under limited data conditions.

\vspace{-0.15cm}
\subsection{Implementation Details}
All models were implemented in PyTorch and trained on an NVIDIA A100 GPU with 80GB VRAM. The dataset was split into 60\% training, 10\% validation, and 30\% testing, ensuring no patient-level leakage between multi-phase and single-phase data. We used a batch size of 8 and trained models for up to 150 epochs with early stopping. To address class imbalance, a balanced sampler was employed to ensure each batch contains both ER and NER cases. The multi-phase and single-phase branches were optimized separately using AdamW with learning rates of 5e-4 and 3e-4, respectively. To mitigate overfitting and unstable updates caused by the limited size of the multi-phase dataset, we adopted a linear warm-up scheduler that gradually increases the learning rate of the multi-phase branch over the first 10 epochs. We set the prototype projection dimension $d$ to 512 in all experiments.

\vspace{-0.15cm}
\subsection{Evaluation Metrics}
\label{sec:eval}
We evaluate model performance using the area under the precision-recall curve (AUPRC), the area under the receiver operating characteristic curve (AUROC), F1 score, sensitivity, and precision. Given the class imbalance in our task, AUPRC is prioritized as the primary metric \cite{saito2015precision}. For threshold-sensitive metrics, we apply two strategies, both using 100 bootstrap iterations: (1) For calculating the F1 score, we adopt the F1-optimal threshold that maximizes F1, providing a balanced and data-driven summary of performance. (2) The second strategy evaluates sensitivity and precision under a fixed specificity of 0.65, reflecting clinically relevant constraints on false positives and offering a conservative yet practical view of recurrence prediction.

\vspace{-0.15cm}
\subsection{Baseline Models}

To evaluate the effectiveness of DuoProto for ER prediction, we compare it against representative baseline models under consistent single-phase input settings.

\textbf{Radiomics model \cite{radiomics}}: Radiomics is a widely used approach for ER prediction in HCC. We extracted handcrafted features from the liver region using the PyRadiomics toolkit and trained an XGBoost classifier. This serves as a conventional benchmark without deep learning.

\textbf{Image-based deep models}: We implemented several SOTA 3D classification architectures commonly adopted in medical imaging. ResNet10~\cite{resnet10} is a convolutional backbone with residual connections. Vision Transformer (ViT)~\cite{vit} applies global self-attention to model long-range dependencies in 3D volumes and has shown promise in medical image classification. Swin Transformer\cite{swin} builds upon ViT by incorporating a hierarchical structure and shifted windows to better capture local context; we initialized it with pretrained Swin-UNETR weights.
We also included ReViT~\cite{ReViT}, a recent SOTA hybrid design that integrates a convolutional mechanism with transformer-based global modeling, which also serves as the feature extractor backbone for DuoProto.

\vspace{-0.2cm}
\section{Results \& Discussions}
\label{sec:result}

\begin{table*}[htbp]
\centering
\caption{Model performance comparison across \textbf{baselines (top)} and \textbf{ablation variants (bottom)}.}

\label{tab:main_results}
\begin{tabular}{lccccc}
\toprule
Method & AUPRC & AUROC & F1 Score & Sensitivity & Precision \\
\midrule
Radiomics \cite{radiomics} & 0.5644 ± 0.2612 & 0.6560 ± 0.1769 & 0.6141 ± 0.2007 & 0.5358 ± 0.1746 & 0.4658 ± 0.1177 \\
Resnet10 \cite{resnet10} & 0.5321 ± 0.1429 & 0.6521 ± 0.1252 & 0.5934 ± 0.1268 & 0.5471 ± 0.1378  & 0.4695 ± 0.0902  \\
ViT \cite{vit} & 0.4758 ± 0.1650 & 0.6307 ± 0.1196 & 0.6355 ± 0.1310 & 0.5214 ± 0.2234 & 0.4526 ± 0.1342 \\
Swin Transformer \cite{swin} & 0.5378 ± 0.1745 & 0.6457 ± 0.1215 & 0.5857 ± 0.1158 & 0.5853 ± 0.1156 & 0.4922 ± 0.0792 \\
ReViT \cite{ReViT} & 0.5647 ± 0.1837 & 0.6657 ± 0.1295 & 0.5817 ± 0.1438 & 0.5786 ± 0.1395  & 0.5013 ± 0.0839 \\
\cmidrule(r){1-6}
w/o $\mathcal{L}_{\text{proto}}$ & 0.5888 ± 0.1828 & 0.6564 ± 0.1388 & 0.6050 ± 0.1217 & 0.5525 ± 0.1027 & 0.496 ± 0.0822 \\
w/o $\mathcal{L}_{\text{sep}}$ & 0.5774 ± 0.1729 & 0.6819 ± 0.1224 & 0.6341 ± 0.1041 & 0.5610 ± 0.1308  & 0.4819 ± 0.0881 \\
w/o $\mathcal{L}_{\text{rank}}$ & 0.5829 ± 0.1777 & 0.6799 ± 0.1245 & 0.6116 ± 0.1187 & 0.5895 ± 0.1624 & 0.4909 ± 0.0835 \\
w/o $\mathcal{L}_{\text{align}}$ & 0.6040 ± 0.1754 & 0.7113 ± 0.1190 & 0.6222 ± 0.1092 & 0.553 ± 0.1316 & 0.4856 ± 0.0817 \\
\textbf{Ours} & \textbf{0.6482 ± 0.1651} & \textbf{0.7438 ± 0.1096} & \textbf{0.6647 ± 0.1204} & \textbf{0.6674 ± 0.1131}  & \textbf{0.5305 ± 0.0739} \\
\bottomrule
\end{tabular}
\vspace{-0.2cm}
\end{table*}

\subsection{Performance Comparison with Baseline Models}

Table~\ref{tab:main_results} (top half) summarizes the performance of baseline models with single-phase PVs. The baselines include a radiomics approach and several SOTA deep learning models designed for medical image analysis.

As shown in Table~\ref{tab:main_results}, our method achieved the best results across all metrics, demonstrating more reliable predictions under clinical application needs. We attribute the superior efficacy to introducing prototype-guided supervision and a dual-branch design, which promotes compact, class-discriminative features. This design helps DuoProto separate subtle recurrence patterns more effectively.

On the other hand, the radiomics model yielded only moderate performance, which may be due to its lack of flexibility to model subtle variations or high-level semantics from imaging data.

Also, unlike DuoProto, classic image-based deep models learn features directly from data, which often lack explicit guidance to focus on recurrence-related cues. Without structured supervision, they may overfit to dominant patterns or background structures, leading to less discriminative representations.

\vspace{-0.15cm}
\subsection{Ablation Study of Prototype-guided Components}

To evaluate the contribution of each component in our framework, we conducted ablation experiments by removing one loss term at a time, as shown in the bottom half of Table~\ref{tab:main_results}.

Removing the contrastive prototype loss ({w/o $\mathcal{L}_{\text{proto}}$}) resulted in a notable $-9.1\%$ drop in AUPRC and $-11.8\%$ in AUROC, along with lower F1 score and sensitivity, indicating that class prototypes are crucial for stabilizing decision boundaries. Excluding the separation loss ({w/o $\mathcal{L}_{\text{sep}}$}) caused a $-10.9\%$ decrease in AUPRC, suggesting that encouraging inter-class repulsion helps preserve decision boundary clarity and prototype distinctiveness.

Excluding the BCLC-informed ranking loss ({w/o $\mathcal{L}{\text{rank}}$}) also degraded AUPRC by $10\%$, confirming that even weak clinical structure offers valuable supervision. Supplementary II further validates the role of BCLC ordering. 

Eliminating the alignment loss ({w/o $\mathcal{L}_{\text{align}}$}) caused a $-6.8\%$ drop in AUPRC, indicating that prototype alignment is essential for maintaining cross-branch consistency. This degradation effectively reduces the model to a single-branch setting. The architectural impact will be further analyzed in the next section.

In summary, the ablation results confirm that each component contributes distinct supervisory signals, and their combination leads to more accurate, stable, and clinically meaningful predictions.
\vspace{-0.15cm}
\subsection{Effect of Dual-Branch Architecture}

\begin{table}[htbp]
\centering
\caption{Comparison between single-branch and dual-branch architectures.}
\label{tab:dual_branch}
\begin{tabular}{lcc}
\toprule
Method & AUPRC & AUROC \\
\midrule
Single-branch (PV) & 0.6040 ± 0.1754 & 0.7113 ± 0.1190 \\
Dual-branch (Ours) & \textbf{0.6482 ± 0.1651} & \textbf{0.7438 ± 0.1096} \\
\bottomrule
\end{tabular}
\vspace{-0.1cm}
\end{table}

To assess the impact of incorporating multi-phase information, we compared the inference efficacy of our dual-branch model against a single-branch variant using only the PV phase for training. As shown in Table~\ref{tab:dual_branch}, the dual-branch design improved AUPRC by 7.3\% and AUROC by 4.6\%. These results suggest that the auxiliary multi-phase branch provides complementary information that enhances recurrence detection. See Supplementary III for more comparisons.
\vspace{-0.15cm}
\subsection{Comparison of Alignment Strategies}

\begin{table}[htbp]
\vspace{-0.2cm}

\centering
\caption{Comparison of cross-branch alignment strategies.}
\label{tab:alignment_strategies}
\begin{tabular}{lcc}
\toprule
Method & AUPRC & AUROC \\
\midrule
Hard-parameter sharing& 0.5366 ± 0.1793 & 0.6611 ± 0.1272 \\
Soft-parameter sharing & 0.5189 ± 0.1557 & 0.6413 ± 0.1210 \\
Prototype alignment (Ours)  & \textbf{0.6482 ± 0.1651} & \textbf{0.7438 ± 0.1096} \\
\bottomrule
\end{tabular}
\vspace{-0.1cm}

\end{table}

We compared three cross-branch alignment strategies: hard parameter sharing via a shared projection head, soft parameter sharing using class-wise feature centroid alignment across the projection head output, and prototype-guided semantic alignment. While both hard and soft sharing promote feature consistency, they lack explicit semantic guidance. In contrast, our prototype-guided design provides structured supervision through learnable class anchors, encouraging discriminative and generalizable representations.

\vspace{-0.2cm}
\subsection{Visualization and Interpretability}

\begin{figure}[!t]
\includegraphics[width=\linewidth]{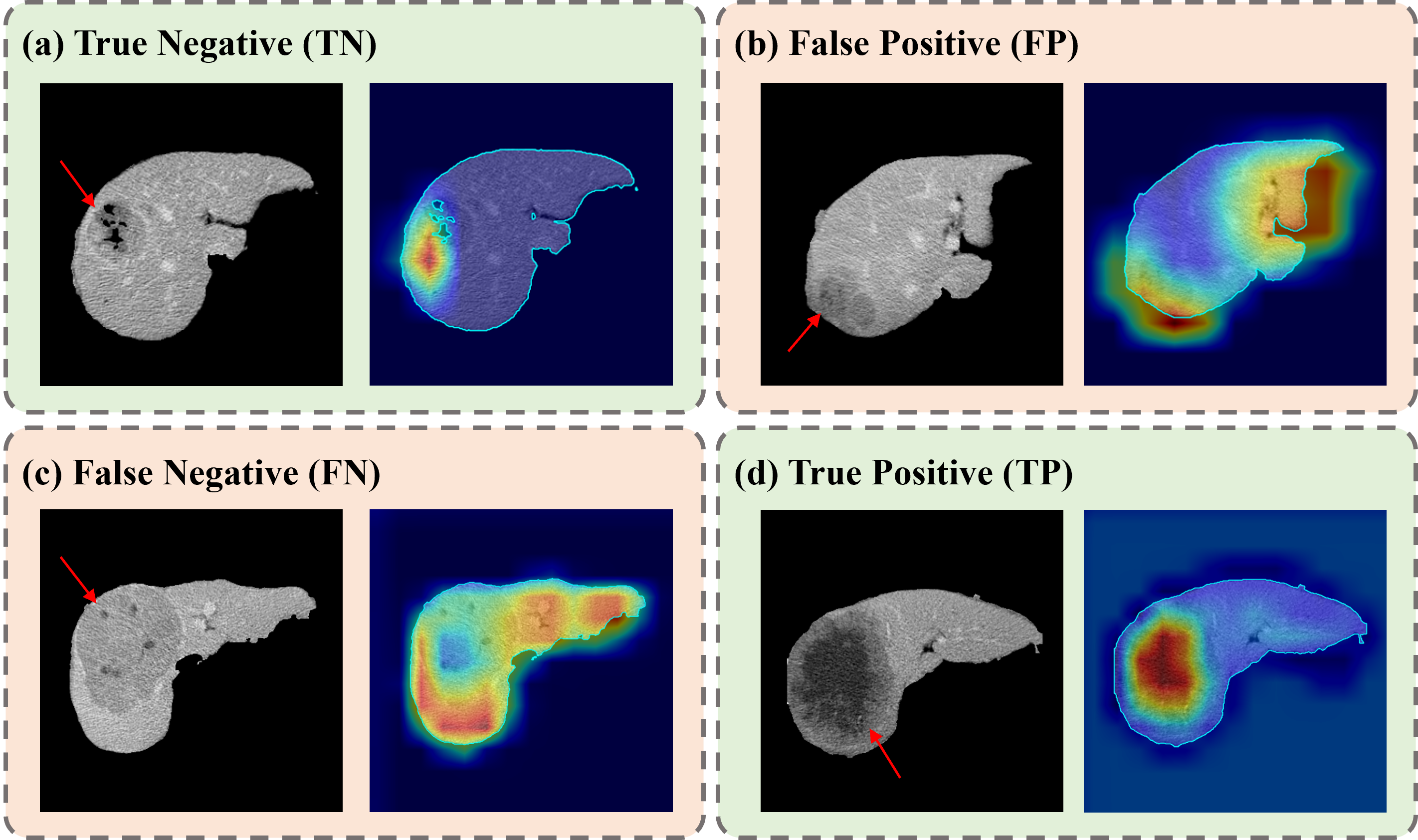}
\caption{Visual explanation of model predictions using Grad-CAM. Each row shows the original CT with liver mask (left) and the corresponding Grad-CAM heatmap (right). Red arrows indicate tumor locations. Green background: correct predictions (TP, TN); red background: incorrect predictions (FP, FN). (a) TN – model correctly predicts non-recurrence despite the presence of a small lesion. (b) FP – misclassifies due to high attention paid to a non-recurrence structure. (c) FN – fails to attend to an evident tumor. (d) TP – correctly focuses on a prominent tumor region.}
\label{fig1}
\vspace{-0.5cm}
\end{figure}

To evaluate the interpretability and localization capability of our model, we generated Grad-CAM heatmaps \cite{selvaraju2017grad} for selected patients across four representative cases, as illustrated in Fig.~\ref{fig1}. These visualizations highlight key regions influencing DuoProto’s predictions. In the true positive (TP) case, the model sharply attends to the tumor region, consistent with clinical expectations. Notably, in the true negative (TN) case, the model correctly predicts non-recurrence despite the presence of a small lesion, suggesting that it captures higher-level patterns such as lesion size or contrast enhancement characteristics. In the false positive (FP) case, the model is misled by a non-recurrent hyperintense region, likely corresponding to benign anatomical structures. In contrast, the false negative (FN) case demonstrates that the model may overlook obvious tumor cues, revealing a sensitivity limitation requiring refinement. These findings not only offer interpretability into the model’s predictions but also underscore areas for future improvement, particularly in enhancing sensitivity to recurrent lesions.

\vspace{-0.3cm}

\subsection{Limitations and Future Work}

One major limitation of our approach is the limited number of available multi-phase samples, which may constrain the stability and generalizability of prototype alignment. Although our joint training strategy helps mitigate this issue to some extent, the model still depends heavily on single-phase supervision.
Additionally, we assessed the prototype-sample distribution using t-SNE visualization \cite{tsne}, which revealed only mild semantic separation in the projected feature space. This observation suggests that further improvement in training convergence and feature discriminability is needed.
As part of future work, we plan to expand the dataset by leveraging additional real-world CT samples from multiple medical institutions through privacy-aware federated learning (FL). This will support both improved training and robust external validation, facilitating broader clinical applicability.

\section{Conclusion}
In this work, we proposed DuoProto, a dual-branch prototype-guided framework for ER prediction in HCC using 3D contrast-enhanced CT. The framework leverages limited multi-phase data as auxiliary input during training to improve learning from widely available single-phase scans, reflecting real-world application needs. By incorporating structured prototype supervision, DuoProto facilitates robust representation learning and effectively bridges discrepancies between imaging phases. Extensive experiments demonstrate that the method consistently outperforms baseline approaches, particularly under class imbalance and missing-phase conditions. Overall, DuoProto offers a practical and robust solution for recurrence risk stratification, supporting more informed clinical decision-making and contributing toward the development of intelligent, data-driven medicine.
\vspace{-0.2cm}

\section*{Ethics Statement}
The in-house dataset used in this study was approved by the Institutional Review Board of National Taiwan University Hospital (IRB No. 202306004RINC). All data collection and analysis procedures complied with relevant ethical guidelines and regulations.

\vspace{-0.3cm}

\section*{Acknowledgement}
This work was sponsored by the National Science and Technology Council (NSTC, 113-2222-E-002-008), the Ministry of Health and Welfare (MOHW, 114-TDU-B-221-144003), and the Ministry of Education (MOE, 113M7054) in Taiwan. Also, this work was financially supported by the “Center for Advanced Computing and Imaging in Biomedicine (NTU-114L900701)” from the Featured Areas Research Center Program within the framework of the Higher Education Sprout Project by the MOE in Taiwan.

\vspace{0.3cm}

\vspace{-0.5cm}
\bibliographystyle{IEEEtran}
\bibliography{references}

\end{document}